# Contrasting Irradiation Behavior of Dual Phases in Ti-6Al-4V Alloy at Low-Temperature Due to ω-phase Precursors in β-phase Matrix

Taku Ishida[1,2*], Sho Kano[3], Eiichi Wakai[1,4], Tamaki Shibayama[5], Shunsuke Makimura[1,2], Hiroaki Abe[3]

[1]    Japan Proton Accelerator Research Complex (J-PARC), Tokai-mura, 319-1195, Japan
[2]    High Energy Accelerator Research Organization (KEK), Tsukuba, 305-0801, Japan
[3]    Nuclear Professional School, School of Engineering, The University of Tokyo, Tokai-mura, 319-1188, Japan
[4]    Japan Atomic Energy Agency (JAEA), Tokai-mura, 319-1106, Japan
[5]    Division of Quantum Science and Engineering, Faculty of Engineering, Centre for Advanced Research of Energy and Materials, Hokkaido University, Sapporo, 0608628, Japan

*    Corresponding author

# Abstract

Aiming to simulate the radiation damage effect on a dual α+β phase Ti-6Al-4V alloy utilized as high-intensity accelerator beam window material, a series of irradiation experiments were conducted with a 2.8 MeV-$Fe^{2+}$ ion beam in several dpa regions at room temperature. The nano-indentation hardness increased steeply at 1 dpa and unchanged up to 10 dpa, due to the saturation of defect clusters and tangled dislocations in the dominant α-phase matrix with a size of 2~3 nm and a density of about $1\times10^{23}$ $m^{-3}$. In contrast in the intergranular β-phase, larger loops of 20~30 nm diameter were observed with much less density of about $5\times10^{20}$ $m^{-3}$. The diffraction pattern showed rectilinear diffuse streaks between the β-phase reflections, corresponding to the ω-phase precursor, without dose dependency in its intensity. FFT/I-FFT analysis of the HREM revealed a sub-nanometer-sized lattice disorder with local fluctuations, not discrete but continuous, and homogeneously distributed within the matrix β-phase stably against the irradiation. The significantly low dislocation density and the absence of phase transformation in the β-phase matrix could be attributed either to the strong sink effect expected for this distinctive sub-nanometer-sized homogeneous lattice disorder or to the anomalous point defect recombination induced by the high mobility of vacancies, both of which are originated from the metastable ω-phase precursors specifically formed in the β(BCC) phase of group-4 transition metals.

# 1. Introduction

Titanium and titanium alloys have principal values for their high strength/weight ratio and good corrosion resistance and are today utilized for a range of industrial fields[1–3]. So far, a wide variety of grades have been developed mainly for aerospace demands[4], which can broadly be classified into three groups based on their phase composition: single α (hexagonal close-packed, HCP) phase, dual α+β phase, and metastable β (body-center cubic, BCC) phase alloys. Recently they have been applied to components in accelerator target systems, such as the production targets and the beam windows (**Fig. 1**a), where heat and thermal shock resistance are particularly required. When these components are subjected to high-energy particle beams, it results in a deterioration of the mechanical properties due to the radiation damage. As nuclear reactor and fusion communities do for a long, it becomes critical for the accelerator community to evaluate the durability of the materials adopted for the apparatus and investigate their radiation damage mechanisms[5,6]. Meanwhile, the investigation into radiation damage effects on titanium alloys, concerning their classifications and microstructural differences, is still at an unsatisfactory stage[7]. This might be because they were not adopted as the materials for the critical components of the fusion reactor, such as first-wall or blanket applications[8], due to their relatively low operating temperature at the time and the potential for high tritium inventory[9].

Among industrial high-strength titanium alloys, Ti-6Al-4V (hereafter referred to as Ti-64) is a standard dual α+β phase alloy[10] and is a commonly used material for beam window applications in accelerator target facilities[11]. It is due to its superior combination of properties against the beam impact: good fatigue endurance limit, low thermal expansion coefficient, and low elastic modulus to reduce stress wave amplitude. It is also because Ti-64 is the only grade of high-strength titanium alloy that has sufficient stock on the market and is available in a variety of product forms. At the neutrino beam-line facilities of the Japan Proton Accelerator Research Complex (J-PARC)[12] and Fermi National Accelerator Laboratory (FNAL)[13], Ti-64 is utilized and to be adopted as the beam windows (**Fig. 1**b) and the target containment vessels and windows[14]. The Large Hadron Collider (LHC) at CERN adopted Ti-64 beam windows for the external beam dump to accommodate its beam power upgrade[15]. At the Facility for Rare Isotope Beams (FRIB)[16], a water-cooled, rotating thin-shell metal drum made of Ti-64 is developed for the beam dump. The International Linear Collider (ILC) is also going to adopt it for the beam window of the main water beam dump[17].

Although Ti-64 is being adopted for critical components at these facilities, it is recognized that the mechanical properties of the alloy degrade immediately by radiation damage. Neutron irradiation in nuclear reactors[18–22] and proton beam irradiation by accelerator[23] have shown that Ti-64 is much more sensitive to radiation damage than single α-phase alloys, such as Ti-5Al-2.5Sn, Ti-6242S, and near-α Ti-3Al-2.5V. Especially at low temperatures, a small amount of irradiation, 0.3 dpa or less, causes significant hardening and embrittlement, resulting in a complete loss of ductility[24]. Formation of defect clusters[18,19,23] and vanadium-rich plate-like precipitates in the α-phase matrix at higher temperature irradiation[18,19] have been reported as its causes, but no conclusive interpretation has been made on the difference in irradiation behavior between Ti-64 and α-alloys at their serviceable low temperature.

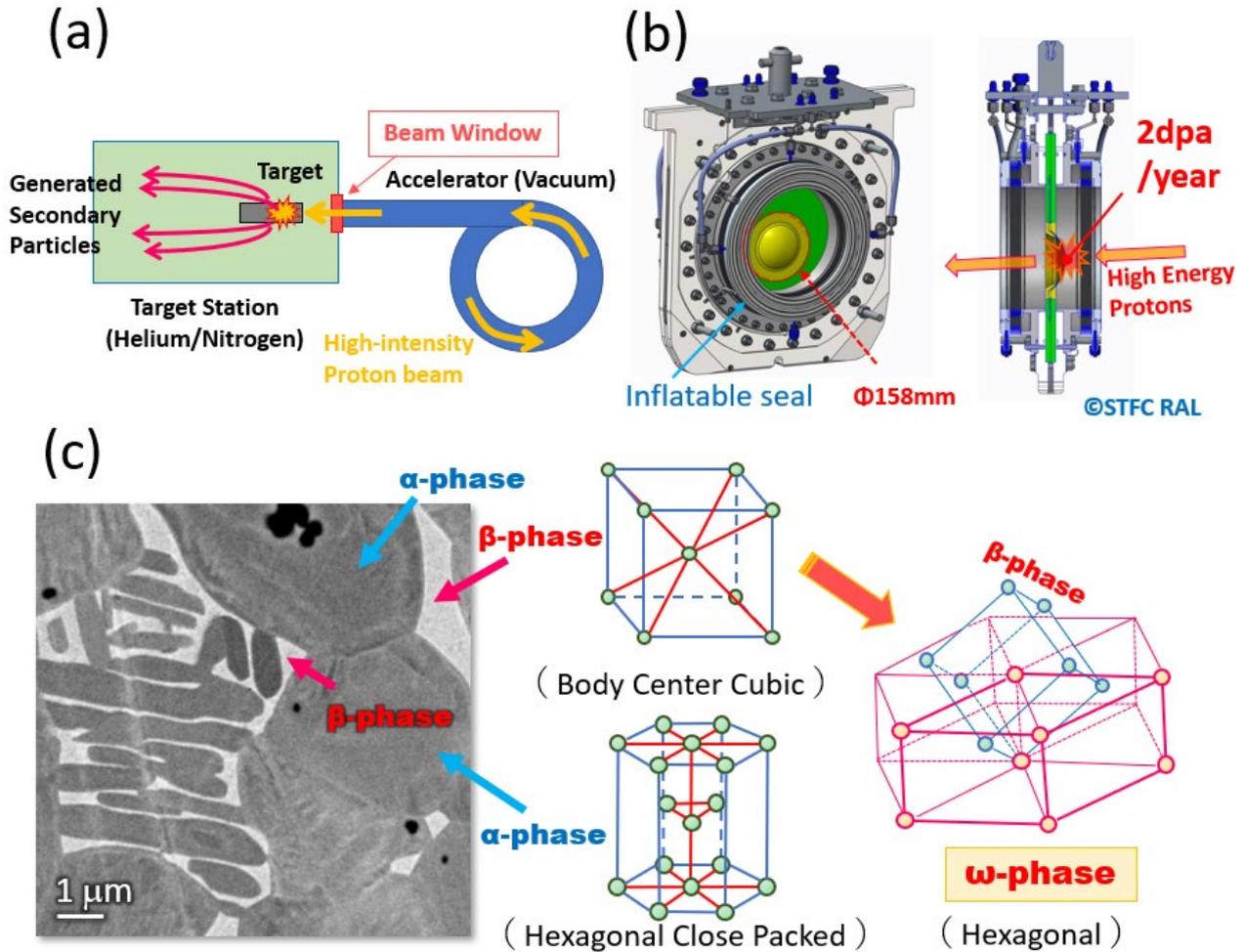

Fig. 1. (a) Conceptual diagram of accelerator beam window. The target that produces the secondary particles is often placed in a target station filled with helium or nitrogen, and the accelerator vacuum and the target station are separated by a thin sheet of metal called a "*beam window*". (b) Beam window at the J-PARC neutrino facility. Helium gas flows through a gap between two 0.4mm thick domed high-strength Ti-64 alloys to cool the heat generated by the beam. It is surrounded by an inflatable seal under pressure for remote handling. The maximum expected radiation damage to the Ti-64 window is approximately 2 dpa/year. (c) The microscope image of Ti-64 shows a mixture of dominant primary α(HCP)-phase and inter-granular β(BCC)-phase matrix. The ω-phase is fine precipitation with a Hexagonal structure in the mother β-phase with coordination relationships $[0001]_\omega // [111]_\beta$, $(11\bar{2}0)_\omega // (110)_\beta$.

When the group-4 transition metal (Ti, Zr, Hf) alloys containing critical β-stabilizing elements are rapidly cooled from the high-temperature β-phase region and subsequently aged at the low to medium temperature (200~400°C), metastable non-close-packed hexagonal precipitates of about a few to several tens of nm, the ω-phase, forms in mother β-phase matrix in high density[25,26] (**Fig. 1**c), which cause the increase of hardness and embrittlement of the mother β-phase matrix, known as the ω-*embrittlement*[27,28]. Historically, they are classified into two categories:

The *athermal* ω phase forms after the quenching as a result of a purely displacive process (the periodic collapse of the $\{111\}_\beta$ planes), which are identified as characteristic diffuse streaks in the diffraction patterns either in $[011]_\beta$ or $[113]_\beta$ zone axes[29]. In contrast, the *isothermal* ω precipitates are produced by the compositional diffusion process activated by the aging heat treatment, which is identified as discrete extra spots at the 1/3 and 2/3 $\{112\}_\beta$ locations in the diffraction patterns[30]. Recent high-resolution electron microscopy (HREM) clarified the formation and development of the ω-phase in atomic-scale lattice image observation[31–33]. In a work utilizing high-angle annular dark field (HAADF)-high resolution scanning transmission electron microscopy (HR-STEM) with atom probe tomography (APT), it is confirmed that the athermal ω phase is accompanied also by compositional fluctuation in a very local scale, and the change from athermal to isothermal ω phase is concurrent[34]. Hereafter, objects with an unstable crystalline structure in the β-phase produced by rapid cooling processes and visible as streaks in diffraction patterns will be referred to as the ω *phase precursor*, and the isothermal ω phase as the *fully-developed ω phase*. Besides these thermal processes, the formation of the ω phase is also known to be induced by deformation, high static or dynamic pressure, or particle irradiation[35–37]. It should be noted that so far the ω-phase is generally investigated in metastable β-phase Ti alloys, and there are very few reports of the ω-phase being observed in dual-phase Ti-64 alloy[38,39]. It has been argued that its formation is suppressed by the presence of sufficient α-phase stabilizing elements (Al and O) in the β-phase[40]. Meanwhile, in a recent high-energy proton irradiation experiment[41], it has been reported that Ti-64 irradiated by the proton beam with 180 MeV kinetic energy up to 0.25 dpa exhibits not only defect clusters in the α-phase matrix in high density but also well-developed ω-phase precipitates in the β-phase matrix with an average size of ~2 nm and a very high number density of $10^{24}$ m$^{-3}$[42]. The greater loss of ductility for the dual α+β phase Ti-64 than that for the single α-phase alloys may reasonably be interpreted as the synergistic effects of the hardening of the α-phase matrix due to dislocations and the ω-embrittlement of the β-phase matrix.

Recent demand for accelerator Ti alloy beam windows is behavior at a much higher irradiation dose. For example, a power upgrade is under progress at the J-PARC neutrino facility[12] with expected annual radiation damage of about 2 dpa to be accumulated on the Ti-64 beam window[11]. However, conducting a proton beam irradiation experiment for material R&D in such a high irradiation dose region is hard to realize. As an alternative approach, ion beam irradiation experiments, whose damage region concentrates at a depth of one to a few micrometers, can accumulate significant radiation damage in a relatively short time[43,44]. It is also advantageous that, unlike proton/neutron irradiation, the irradiated specimens do not significantly become radioactive, and the irradiated specimens can be handled in ordinary laboratories. The nano-indentation can provide hardness, which is about in proportion to the macro-scale mechanical strength[45,46]. So far ion beam irradiation studies on Ti-64 were made at high temperatures, typically at several 100 °C regions with emphasis on the precipitation in the α-phase matrix aiming at nuclear and fusion reactor applications[47–54]. Meanwhile, the serviceable temperature of Ti-64 as a structural material is limited to less than 300~400 °C[55], and the Ti-64 beam window with our interest is operated typically with forced cooling (such as Helium gas or water) to keep its temperature well under the serviceable temperature limit, at most 200~250°C.

In the present study, a 2.8 MeV Fe$^{2+}$ ion beam irradiated Ti-64 specimens up to 11 dpa at room temperature, which is comparable to the radiation damage accumulated on the beam windows at the relevant proton accelerator facilities under several years of operation. Transmission electron microscopy (TEM) has been conducted to investigate the microstructural evolutions of each phase, and HREM analysis has been performed on the β-phase to investigate dose dependency of the ω-phase manifestation in more detail. The goal of the study is to evaluate the irradiation hardening of Ti-64 and confirm the material integrity under high irradiation dose region comparable to the future accelerator beam windows, and to correlate the hardening behavior to the microstructural changes in the α and β phases. Here, special attention is imposed on evaluating the behavior of the radiation-induced ω phase precipitation in the β-phase, which can be the critical controlling factor for irradiation embrittlement of Ti-64.

| Al | V | O | Fe | C | N | H | Ti |
|---|---|---|---|---|---|---|---|
| 6.00 | 4.07 | 0.112 | 0.16 | 0.023 | 0.006 | 0.0054 | Bal. |

Table 1 Chemical composition (wt. %) of the Ti-64 sheet material provided for current studies.

## 2. Material and methods

In this study, a 1 mm-thick sheet of Ti-64 Extra-Low Interstitial (ELI) alloy, a product of VSMPO-AVISMA Co., was purchased from the stock market, whose chemical composition is listed in **Table 1**. The material in mill-annealed condition (1,063 K for 3.06k sec, Air-Cooled) was stress-relief annealed (868 K for 10.8k sec) and then furnace-cooled with argon gas flow. Disc specimens with a size of 3 mm $\phi$ ×0.2 mm thick were prepared, whose surface was mirror-finished by polishing with alumina slurry and colloidal silica suspensions. The deformed layer induced by the polishing was removed by the electrochemical etching in a solution of 6% perchloric acid, 35% butanol, and 59% methanol at 235 K, with a voltage of 19 V, a current of 34 mA, and polishing time of 20 sec. Prior to the irradiation, the morphology of the material was investigated by X-ray diffraction (XRD) by laboratory Cu-Kα X-ray equipment with a power of 40 kV×40 mA, and by electron backscatter diffraction (EBSD) with the acceleration voltage and step size set to 30 kV and 0.2 µm, respectively. The chemical composition of each phase was investigated by SEM-Energy Dispersive X-ray Spectroscopy (EDS).

The ion beam irradiation was conducted at the High Fluence Irradiation Facility, University of Tokyo (HIT)[56], where the $Fe^{2+}$ ion was accelerated by an HVEE Tandetron Accelerator to 2.8 MeV kinetic energy and irradiated the specimens at room temperature (RT, ~298 K). The depth profile of radiation damage was estimated as shown in **Fig. 2** by the Stopping and Range of Ions in Matter (SRIM)[57] with the full cascade option, where the displacement threshold energies for Ti, Al, and V were set to 30, 27, and 57 eV, respectively[58]. The damage peak is located at a depth of about 1.2 µm from the surface of the specimen, where the damage rate is estimated to be about $1\times10^{-3}$ dpa/sec. The $Fe^{2+}$ ion implantation exhibited its peak at a depth of 1.45 µm from the specimen surface. At the damage peak in the specimen irradiated to 11 dpa, the concentration of $Fe^{2+}$ ion is estimated to be 0.32 wt.% (0.27 at.%), which is comparable to the typical impurity of the standard grade of Ti-64 and the effect on the hardness is considered to be limited. The duration of the beam exposure was controlled to accumulate radiation damage of 1, 5, and 11 dpa at the damage peak. The temperature of the specimen was monitored by multiple thermocouples installed on each specimen holder, which confirmed no beam heating occurred during the irradiation, because of a gold foil between the specimens and the holder for good thermal contact.

Nano-indentation hardness measurements were performed on specimens irradiated at different dose levels using a Berkovich-type indenter. The depth of the indentation was set to 150 nm by referring to the estimated damage profile described above. The indentation was repeated 120 times for each specimen at 10 µm intervals, and hardness was deduced using the technique detailed by Oliver and Pharr[59], where ineligible data were manually

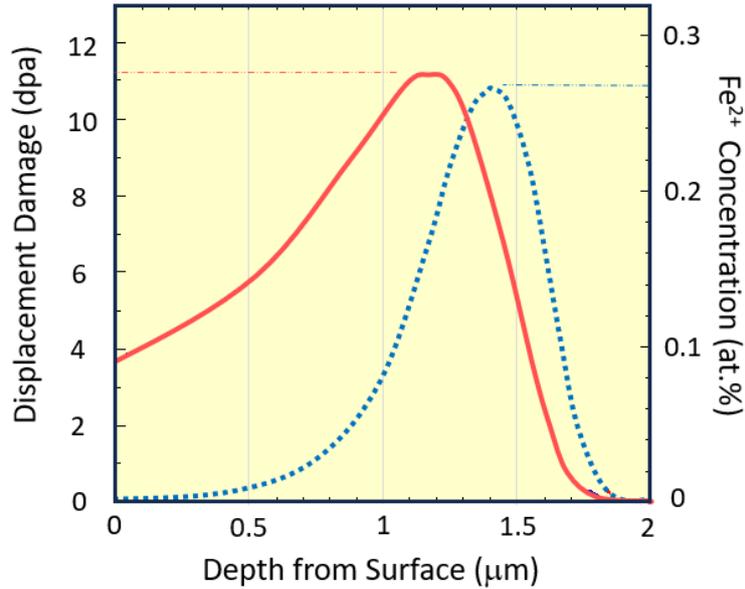

**Fig. 2.** Accumulated displacement damage and $Fe^{2+}$ concentration as functions of depth from the surface of the Ti-64 specimen, estimated by SRIM code.

excluded based on indentation depth and load curve behavior.

For TEM observations, thin lamellae were produced from the disc specimen irradiated to 11 dpa by focused ion beam (FIB) milling according to the in-situ lift-out method[60]. To protect the workpiece from damage by the ion beam, a tungsten deposit layer covering a 25 μm × 2 μm area was prepared in prior. FIB milling was then performed by the Ga ions with the acceleration voltage set to 40 kV. The produced lamellae of less than 300 nm thickness were picked up and mounted on a copper mesh. To remove artifacts caused by the ion beam, the lamellae were flash polished with the electrolyte utilized for the TEM disc polishing, cooled to 233 K for a voltage of 20~25 V and a duration of 150 msec.

TEM observations were made to characterize the microstructural changes of each α and β phase matrix induced by the irradiation. For α(HCP)-phase matrix, two-beam diffraction conditions with $\mathbf{g} = (10\bar{1}0)$ and $(0002)$ were employed to distinguish *a*- and *c*-dislocations following the $\mathbf{g} \cdot \mathbf{b}$ criterion (**b**: Burgers vector). The β(BCC)-phase matrix was investigated with $\mathbf{g} = (200)$ and $(01\bar{1})$ to differentiate dislocations with $\mathbf{b} = 1/2 \langle 111 \rangle$ and $\langle 100 \rangle$. To estimate the number densities and average sizes, multiple defect clusters and loops were examined for each image, where the thickness of the specimen was evaluated by counting the equal-thickness fringes. The formation of the ω-phase in the β-phase matrix was examined from the direction parallel to $Z = [011]_\beta$ zone axis, where the dose dependence of the ω-phase formation was investigated by obtaining diffraction patterns (DP) at different depths. To identify the manifestation of the ω-phase, HREM lattice image analysis of the β-phase matrix was also carried out at the High-Voltage Electron Microscope Laboratory of Hokkaido University, using an FEI Titan3 G2 60-300 operating at 300kV.

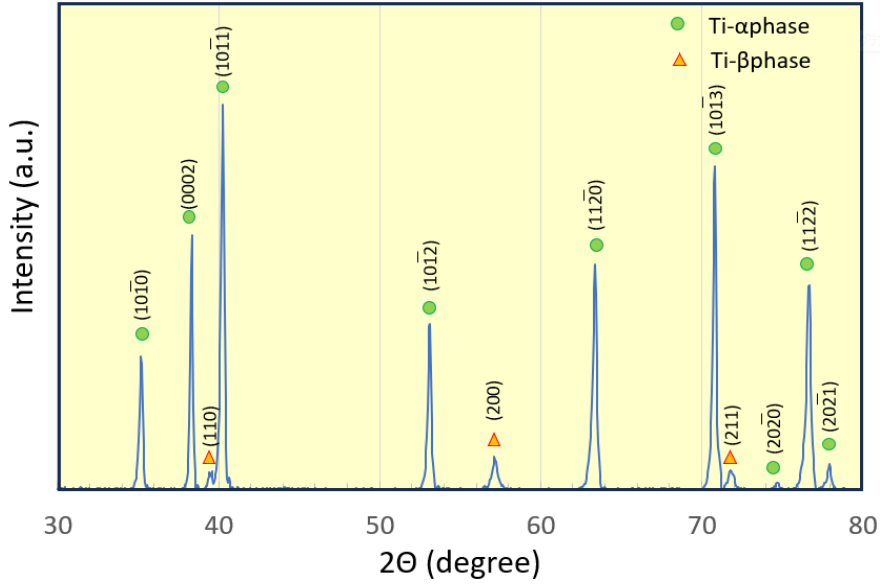

Fig. 3. XRD patterns of a bulk specimen of Ti-64 in un-irradiated condition.

## 3. Results and discussion

### 3.1 Microstructure, phase analysis, and elemental mapping on un-irradiated specimen

**Fig. 3** shows the XRD pattern of the un-irradiated Ti-64 specimen as a bulk, where all observed peaks can be classified into either α or β phase diffractions, and no peaks accompanied by other phases or compounds, such as titanium oxide, were detected. By utilizing the highest peak intensities of each phase, $(10\bar{1}1)_\alpha$ and $(200)_\beta$, and estimating their ideal ratio assuming powder diffraction by CrystalDiffract software[61], the volume fraction of each phase was estimated to be α : β = 78% : 22%. The phase composition of Ti-64 can be controlled by the final heat treatment temperature and cooling speed[62]. The obtained ratio is comparable to the equilibrium at its mill-annealing temperature (790°C) with a fast cooling rate anticipated for the thin sheet.

**Fig. 4** shows the EBSD micrographs of the un-irradiated Ti-64 specimen in parallel to the surface of the sheet material. The primary globular α-phase grains with less than 10 μm in size are dominant, with crystalline planes preferentially oriented to either $(0001)_\alpha$ or $(2\bar{1}\bar{1}0)_\alpha$. The β-phase matrix is distributed at the grain boundaries of the α-phase grains, with major orientation to $(111)_\beta$ plane, as expected by Burger's directional relationship from BCC to HCP transition: $(0001)_\alpha//(110)_\beta$, $[2\bar{1}\bar{1}0]_\alpha//[111]_\beta$[63]. The observed fine microstructure with a moderate α-phase basal texture appears to be typical of the hot-rolled Ti-64 sheet and strip products[55]. The volume fraction of the β-phase was about 10 % in (d) the phase map. The difference with the XRD results (22%) could be due to the limited accuracy of XRD for the textured bulk specimen since specific diffraction intensities are stronger due to the anisotropy. It should also be noted that EBSD observes area fraction at the surface by low-energy backscattered electrons, whereas XRD observes volume fraction to a deeper region through characteristic X-rays.

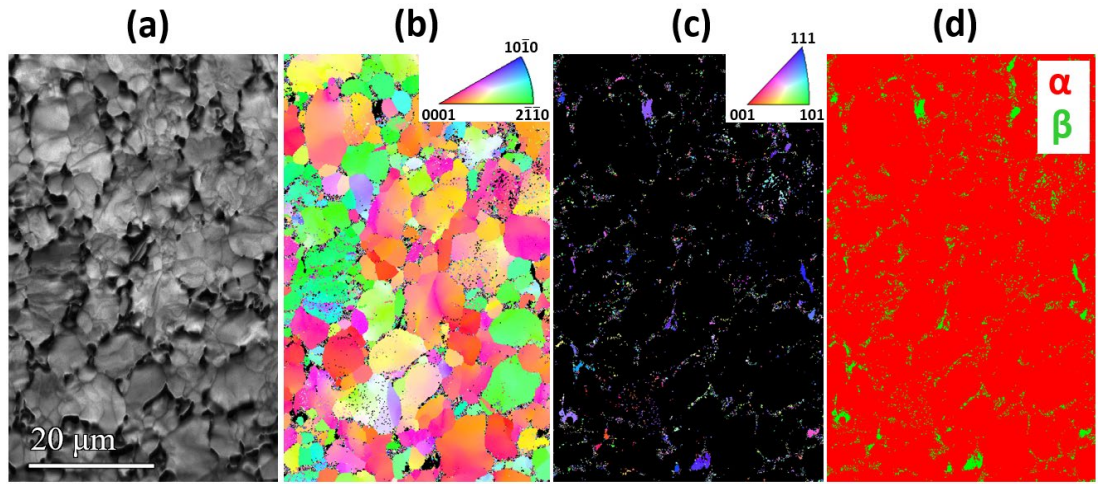

Fig. 4. The EBSD micrographs of the Ti-64 specimen. (a) The image quality (IQ) map, inverse pole figures (IPF) of (b) α and (c) β phases, and (d) phase map.

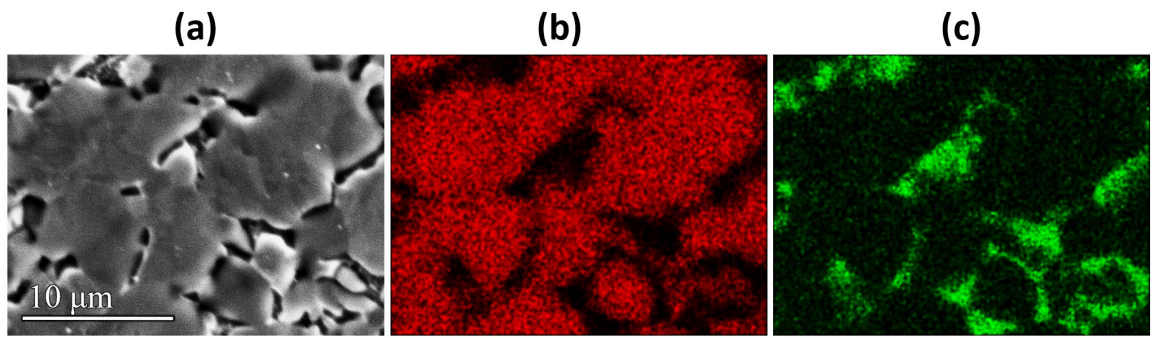

Fig. 5. Micrographs of SEM-EDS analysis of Ti-64 specimen: (a) Secondary Electron Image, (b) Aluminum distribution map and (c) Vanadium distribution map.

The elemental mapping by SEM-EDS is shown in **Fig. 5**, where (a) the secondary electron (SE) image shows an equivalent microstructure as of the EBSD (**Fig. 4**a). The elemental mapping verified an enrichment of (b) aluminum (α-stabilizer) in the α- phase matrices and (c) vanadium (β-stabilizer) in the β-phase matrices, respectively. Since the volume fraction of the β-phase is considerably low, a high concentration of vanadium, more than 10%, is expected in the β-phase[62].

### 3.2 Nano-hardness measurements

The nano-hardness distributions of the Ti-64 specimens irradiated with ion beams under RT to different dose levels are shown in **Fig. 6**. Each distribution exhibited a large deviation, which can be attributed to the anisotropy of the indentation response of the dominant α(HCP)-matrix with limited slip systems. Since the prismatic $a$-glide is the main deformation mechanism for α-titanium, it has the hexagonal $c$-axis as the hardest direction, and the nano-hardness of the basal plane is twice greater than those perpendicular to the prismatic planes[64].

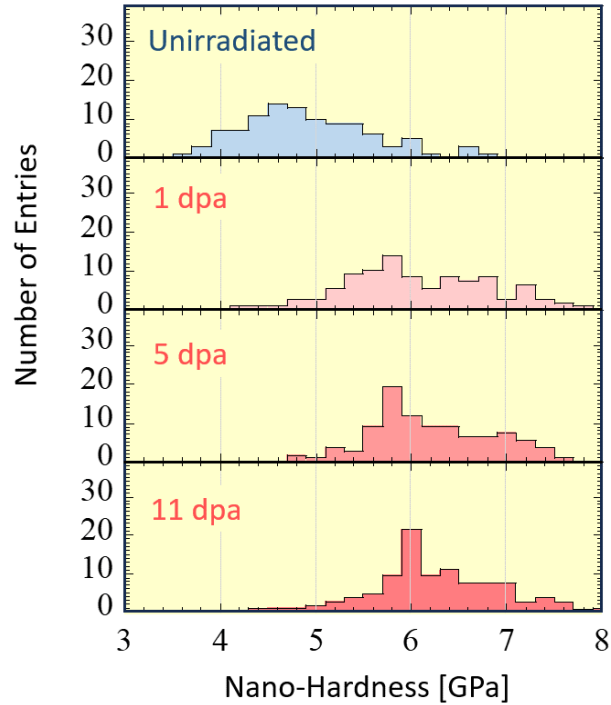

Fig. 6. Nano-hardness distribution of Ti-64 irradiated with 2.8 MeV $Fe^{2+}$ ion beams up to different dose levels under RT.

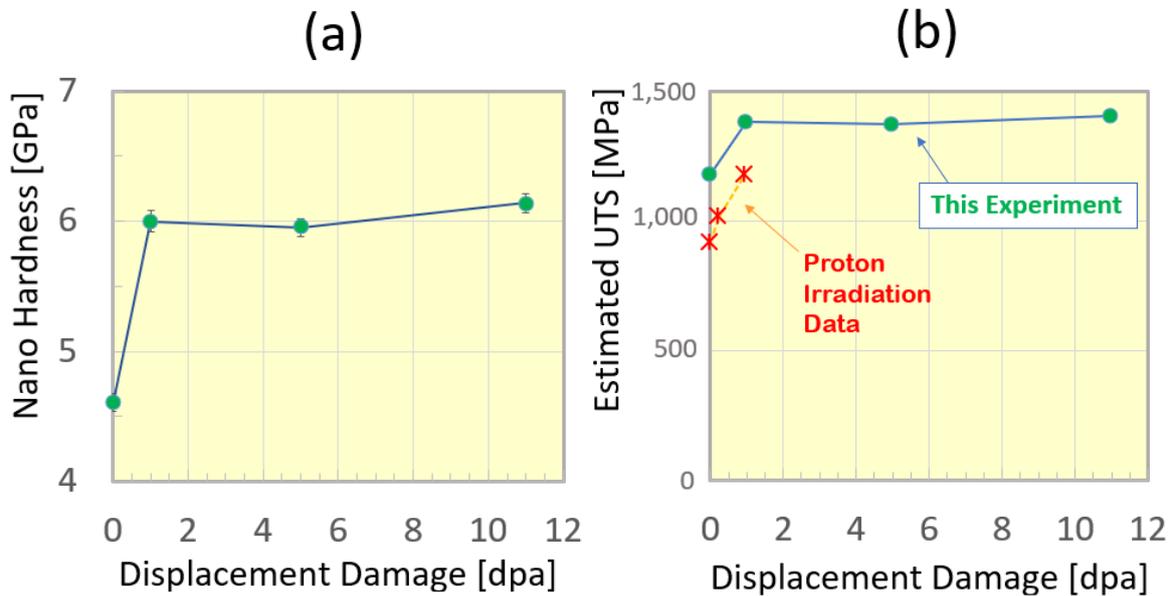

Fig. 7. (a) The average of nano-hardness of Ti-64 as a function of displacement damage by ion beam irradiation under RT. (b) UTS estimated from the nano-hardness by applying an empirical formula. For comparison, tensile data of proton beam irradiation on the same Ti-64 material were overlayed.

**Fig. 7**a plotted the average of the nano-hardness distributions as a function of irradiation dose. It shows that Ti-64 hardens rapidly at the early stage of irradiation to 1 dpa, after which the hardness saturates and stays almost the same value up to 11 dpa. **Table 2** lists the average values of the nano-hardness at each dose, where the increase of the hardness is about equal at 1~11 dpa, 1.4 to 1.6 GPa, in comparison to un-irradiated state. As a reference, the HCP matrix orientational dependence on the radiation hardening was studied for a polycrystalline zirconium alloy irradiated by the self $Zr^{2+}$ ion at 573 K up to 0.2 dpa[65], where the amount of the hardness increment (0.7~0.8 GPa) was observed to be almost uniform and irrelevant to the rotation angle from the *c*-axis. Since the Ti and Zr have about the same *c/a* ratio, the impact of the texture on radiation hardening is also considered to be limited.

It is widely recognized that the hardness and mechanical strength of steels and other common alloy systems obey the empirical *"three-times"* relationship, $H_V \approx 3 \cdot \sigma_y$ or $H_V \approx 3 \cdot \sigma_{UTS}$, where $H_V$ denotes Vickers hardness, $\sigma_y$, $\sigma_{UTS}$ are yield strength, and ultimate tensile strength, respectively[45,46,66]. However similar relationships are not as common for titanium alloys, because the hardness is dependent on the orientation of the HCP matrix as described above. An alternative empirical formula was obtained for investment cast Ti-64 components[67,68]:

$$\sigma_{UTS}[MPa] = \frac{H_V[MPa]}{6.33} + 503.$$

By employing the relation between $H_V$ and nano-hardness $H_n$, $H_V[kgf/mm^2] = 94.495 \cdot H_n[GPa]$[69], $\sigma_{UTS}$ can be estimated from the average of $H_n$, as shown in **Fig. 7**b and listed in **Table 2**. For comparison, macroscale tensile test results are also overlaid in **Fig.7**b, where the tensile specimens were produced from the same Ti-64 sheet material and irradiated by an accelerator proton beam with 180 MeV kinetic energy at low temperature up to ~1 dpa[42,70]. As can be seen, the increase of the strength from the un-irradiated state to 1 dpa, about 200 MPa, is consistent between current ion beam irradiation results and proton beam irradiation measurements. The higher strength values estimated from the nano-hardness than those by macro-scale tensile testing may be due to the texture of the sheet material, as shown in **Fig.4**b: the nano-hardness was dominantly measured perpendicular to the hardest basal plane, whereas the tensile dog-bone specimens are taken parallel to the softer *a*-axis direction.

To differentiate the irradiation hardening of β phase from the dominant α phase matrix, the specimens after nano-indentation measurement were subjected to SEM-EDS analysis. As shown in **Fig. 8**a, the SE image clearly shows indentations regularly aligned at 10 μm intervals; by superimposing the SE image and the EDS elemental mapping, each indent can be classified into two groups, one in the Al-rich region (red) and the others in V-rich region (green). The Al-rich and V-rich regions are thought to consist mainly of the α- and of the β-phase, respectively. **Fig. 8**b shows the nano-hardness distribution for each region of the specimen irradiated to 11 dpa, where the hardness of both regions showed similar broad distribution, likely due to the orientation dependence of the α(HCP) phase hardness, with a somewhat lower mean value for the V-rich region. The similarity of the distribution and the low volume fraction of the β-phase of this material, the hardness distribution of the V-rich region is considered to be not only from the β-phase but affected by the surrounding α-phase matrix and also platelet secondary α-phases formed in the β-phase. **Fig. 8**c exhibits the average value of the hardness for each region as a function of dpa, which shows that the hardness of the V-rich region was significantly lower than that of the Al-rich region at each irradiation dose, which may suggest that the hardness of the β-phase is significantly lower than the α-phase at each irradiation dose. In a past study on proton irradiated β-titanium alloy to 0.06~0.12 dpa at ~100°C, the change of micro-Vickers hardness was found to be negligible[71]. However current analysis is not accurate enough to differentiate the exact amount of the β-phase hardening.

| Irrad.T | Dose | Nano-Hardness [GPa] | Estimated UTS [MPa] |
|---|---|---|---|
| | Unirr. | 4.6±0.7 | 1,180 |
| 298 K (RT) | 1 dpa | 6.0±0.8 | 1,380 |
| | 5 dpa | 6.0±0.7 | 1,370 |
| | 11 dpa | 6.2±0.7 | 1,400 |

Table 2 Average of the nano-hardness distribution and UTS value estimated from the hardness for each irradiation dose. The errors of the hardness indicate the standard deviation of the hardness distribution.

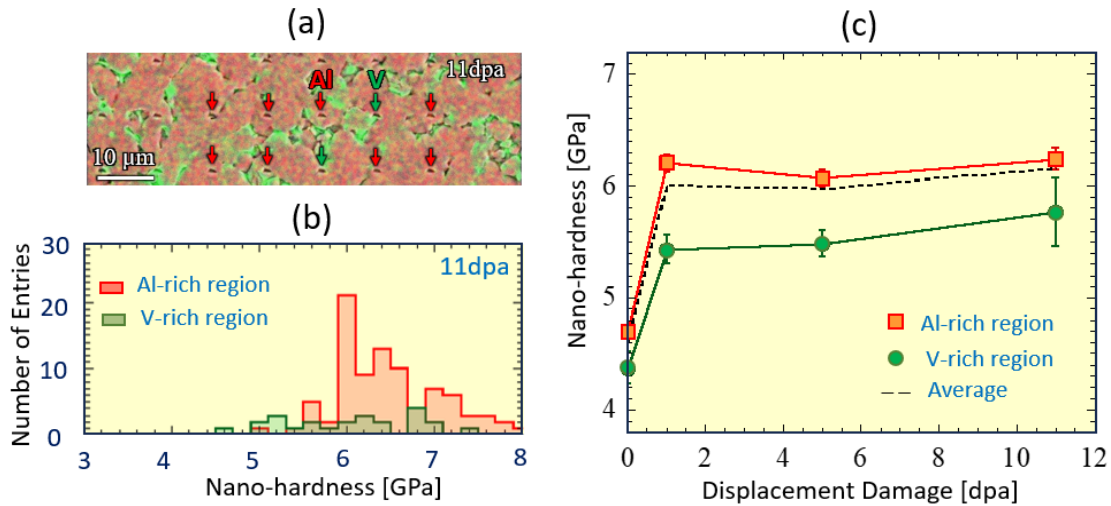

Fig. 8. (a) SEM image of the specimen surface after nano-indentation, overlayed with EDS elemental mapping. Each indent with 10 mm intervals can be classified into one in Aluminum segregation (red) and the others in Vanadium segregation (green). (b) The nano-hardness distribution for each region, irradiated to 11 dpa under RT. (c) The average of nano-hardness for each region as a function of dpa. Note that V-rich region is not only β-phase, but affected by surrounding and/or intergranular α-phase.

## 3.3 TEM and HREM analysis on the irradiated specimen

To characterize the radiation damage microstructure, TEM observations of the Ti-64 specimen irradiated to 11 dpa at the damage peak were carried out.

### 3.3.1 TEM observation on damage morphology in the α and β matrices

It is widely understood that amount of the displacement damage caused by the ion beam is dependent on the depth from the incident surface of the irradiated material. Consequently, the TEM lamella specimen prepared by FIB milling in a direction perpendicular to the material surface can effectively reveal the dpa dependence of the radiation-induced defect morphology. **Fig.9** shows TEM bright-field images of (a) an α grain and (e) a β grain within the irradiated area and their magnified views at different depths. Though the electropolishing after FIB milling resulted in loss of the surface region, the location of the original incident surface can be determined from the remaining protective deposit layer at both ends of the lamella (outside of these figures), which was indicated by blue lines in (a) and (e) (Note that for β-phase the surface may locate left side, since β-phase was electropolished more rapidly than α-phase through TEM disk preparation). As also shown by the yellow lines in these figures, the examined α-grain spans a depth of 0.4~1.0 μm and the β-grain spans a depth from 0.5 μm to more than 2.0 μm from the incident surface. According to the SRIM estimations (**Fig.2**), each range of depth corresponds to approximately 5 ~ 10 dpa for the α-grain, and 6 ~ 11 dpa for the β-grain which covers the entire damage region, respectively. Though ranges of damage amount are comparable, the radiation damage morphology and its depth dependence appear very differently between the two phases: In the α-phase, as shown in **Fig.** (b), (c) and (d), high-density defect clusters/tiny tangled dislocation loops are distributed uniformly, whose small size and high density are rather irrelevant to the depth from incident surface. On the other hand, for the β-phase, high-density tiny defect clusters as observed in α-phase are not visible at all in the depth less than 1 μm (**Fig.** (f)). Instead, weak contrast corresponding to larger dislocation loops are formed in low-density, which turns into a band of overlapping dislocation loops of 20~30 nm size parallel to the incident surface ranging in a 1.0~1.5 μm depth (**Fig.** (g)). The formation of the damage band consisting of these relatively large dislocation loops is depth-dependent, whose damage threshold is considered to be about 10 dpa. In the region with damage less than 10 dpa, the β-phase seems to preserve superior radiation damage resistance than the α-phase.

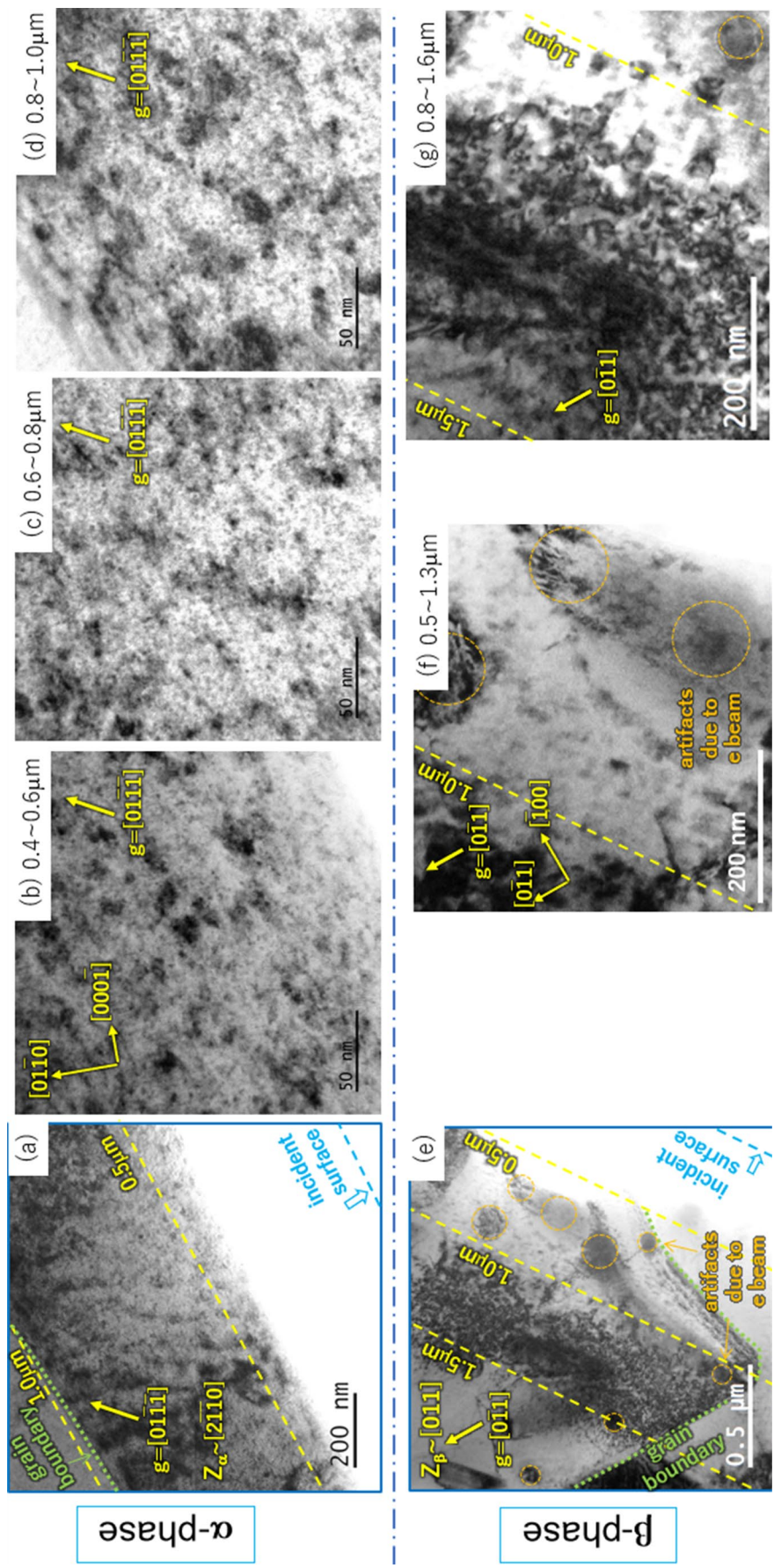

Fig. 9. TEM Bright Field images of the α/β-phase grains in two-beam diffraction condition and magnified views at different depth ranges. (a-d) α-phase views from $Z \approx [2\bar{1}\bar{1}0]_\alpha$ zone axis with $g = [01\bar{1}\bar{1}]$. (e-g) β-phase views from $Z \approx [011]_\beta$ zone axis with $g = [0\bar{1}1]$. In low magnification views (a)(e) incident surface, depth from the surface, and grain boundary are shown by blue, yellow, and green lines, respectively. Note that in this β-phase grain, multiple artifact damages exist at shallow region and grain boundaries, which were caused by prior observations utilizing high-energy electron beam (indicated by orange circles).

**Fig. 10** shows an area of the α-phase matrix at a depth of 0.7 *μm* from the incident surface, whose damage corresponds to 7 dpa. **Fig. 10**a shows the selected area diffraction pattern (SADP), the bright field (BF), and weak beam dark field (WBDF) images of the α-phase matrix in two-beam diffraction condition with an inverse lattice vector $\mathbf{g} = (10\bar{1}0)$. As can be seen, numerous black dot contrasts in BF and corresponding white dot contrasts in WBDF were identified in a very high density. They are the defect clusters and small tangled dislocation loops, that are too tiny to resolve in the conventional TEM image. Although complete characterization of these tiny tangled defects/loops is difficult, they are considered mostly to be the *a*-type perfect loops in the prismatic planes with Burgers vector $\mathbf{b} = a/3 \langle 11\bar{2}0 \rangle$, *i.e.* the shortest lattice vector in the basal plane. **Fig. 10**b is another set of images employing diffraction with $\mathbf{g} = (0002)$, where the loops with *c*-component were identified with similar size to the *a*-loop with somewhat lower density. Again, characterization of these tiny loops is difficult, they may be classified into the partial *c*-dislocation loops in the basal plane with $\mathbf{b} = 1/6 \langle \bar{2}203 \rangle = c/2 \langle 0001 \rangle + a/3 \langle \bar{1}100 \rangle$, *i.e.* the smallest interatomic vector between two neighboring (0001) planes, and occasionally referred to as *c/2+p* loops.

The observed "*black-dot*" damages are typical for irradiation in the "*low-temperature regime*" [72,73], where the mobility of vacancy point defects is so limited that defect clusters and small loop damages are simply accumulated until their saturation. The fact that the nano-hardness of Ti-64 shows a sharp increase at about 1 dpa and then shows almost no change up to 11 dpa (as shown in **Fig.7**) may be due to the saturation of the defect cluster formation in the dominant α-phase. Similar uniformly distributed small defect clusters can be found regardless of FCC/BCC/HCP systems for neutron irradiation at low-temperature[74]. For Ti-64, a similar damage signature was reported in the α-phase for irradiation by neutrons at 50°C[19] and also by protons less than 120 °C[42]. Radiation damage in HCP metals varies from metal to metal, reflecting the relative packing density of different planes, which depends on the *c/a* ratio[75]. Ti and Zr are both group-4 transition metals with very similar *c/a* values, 1.587(Ti) and 1.593(Zr), which is less than the value for ideal HCP crystal, $(8/3)^{1/2} = 1.633$. For α-titanium and zirconium alloys irradiated at the higher temperature (> 300°C), the principal defects are the *a*-type perfect loops in the prismatic planes, which have an elliptical shape with a major axis of 5~10 nm long parallel to the *c*-axis, aligned in rows parallel to the basal planes[76–78]. On the other hand, the *c*-component loops in the basal plane have invariably vacancy nature[75], which only form at the higher dose with their size greater than 50 nm. The formation of large planar *c*-loops is known to cause anisotropic irradiation-induced growth of Zr alloys in the breakaway period, due to a decrease in *c/a* ratio[79,80]. There is an observation that the alignment of *a*-loops may cause nucleation of the *c*-loops[81]. In the current study, however, for the ion beam irradiation of Ti-64 to several dpa at RT, the defect clusters/small dislocations with *c*-component form as comparable (a few nm) size as those for the *a*-loops, which is in contrast to the neutron irradiation at higher temperature. For the high-temperature neutron/ion irradiation on Ti-64, the formation of vanadium-rich β-phase precipitate in the α-phase matrix has been studied intensively as a unique phenomenon for this dual-phase alloy containing significant vanadium content[18,52,54]. In the current study, however, we did not impose much care on this phenomenon, since at RT, mobility of vacancies and resulting diffusion of solute vanadium should be limited in the α-phase, and actually, the precipitate was not observed for Ti-64 irradiated by neutron at 50°C[19].

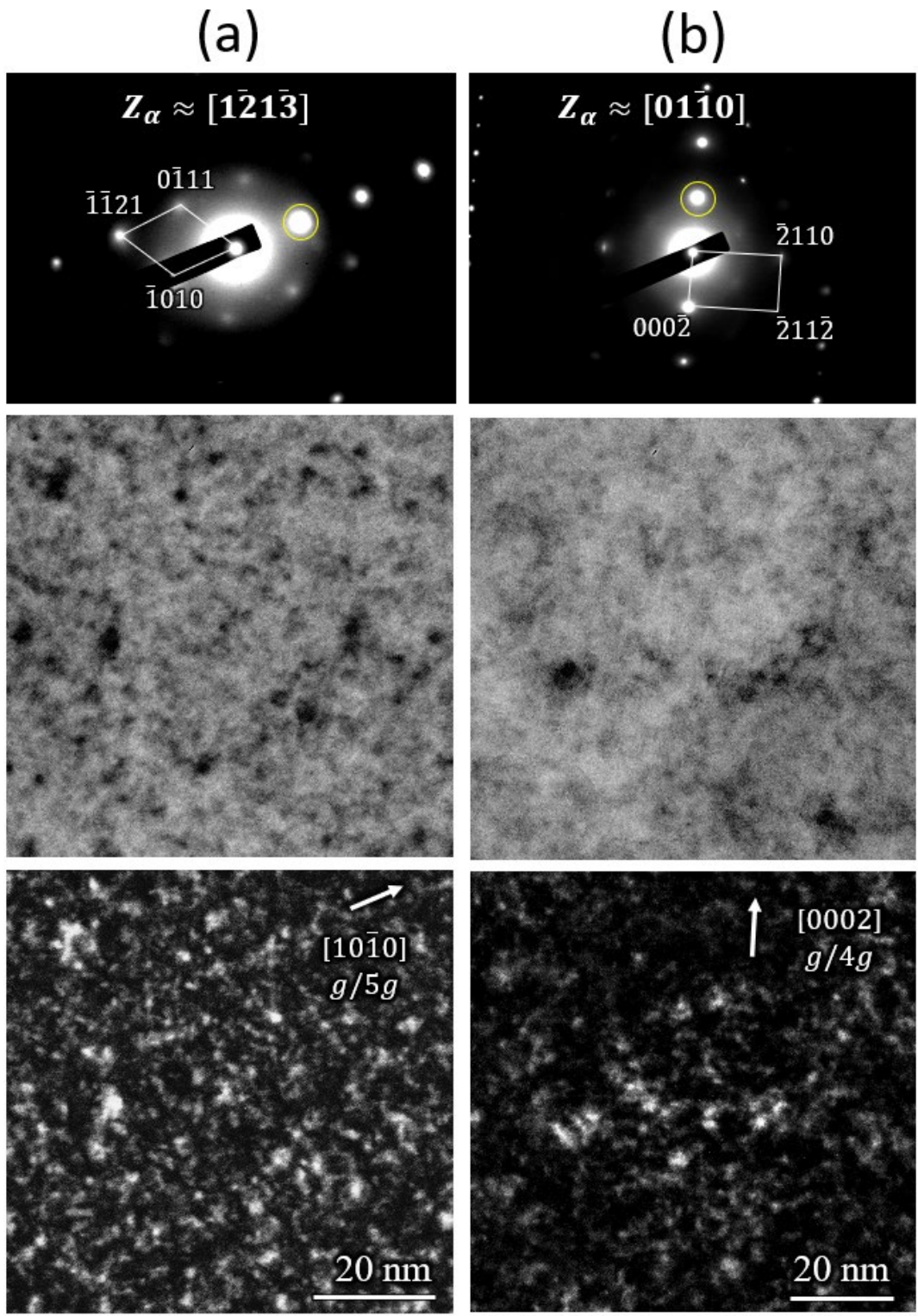

Fig. 10. SADP (top), BF (middle), and WBDF (bottom) images of the α-phase matrix in the area with an accumulated dose of 7 dpa under RT (a) with $Z \approx [1\bar{2}1\bar{3}]_\alpha$ and (b) $[01\bar{1}0]_\alpha$ zone axes. The WBDF images were taken by selecting a diffraction spot indicated by a yellow cycle in each DP.

**Fig. 11** shows sets of SADP, BF and WBDF images on an irradiated β-phase matrix from $Z_\beta \approx [011]$ zone axis, by exciting diffraction for (a) $\mathbf{g} = (200)$ and (b) $\mathbf{g} = (01\bar{1})$, respectively. The figures show a region at a depth range of 0.7~1.2 μm corresponding to 7~11dpa damage which increases from the bottom-right to the top-left, with the most damaged top-left part resting on the band of dislocation loops observed in **Fig. 9**g. As can be seen, the *black-dot* contrasts, *i.e.* high-density defect clusters and small loops which were the dominant damage signature for the α-phase (**Fig. 10**), are not present at all in the entire region. Instead, larger dislocation loops of about 20~30 nm in size can be identified with much lower number density in the shallow area (bottom-left to center, corresponding to damage less than ~10dpa), which rapidly increase and form a band of overlapping dislocation loops at top-right (damage peak, ~11dpa). They show circular/elliptical shapes in (a) and nearly edge-on shapes in (b), respectively. For the BCC metals, there are two types of dislocation loops with $\mathbf{b} = 1/2\langle 111 \rangle$ and $\langle 100 \rangle$. The observed (a) circular/elliptical loops and (b) edge-on loops exhibit comparable sizes and number densities, which are considered to be $1/2\langle 111 \rangle$ loops. It seems to be no $\langle 100 \rangle$ loops were formed, because from $Z_\beta = [011]$ zone axis, loops with $\mathbf{b} = [100]$ should appear as edge-on in (a) $\mathbf{g} = (200)$, and loops with $\mathbf{b} = [010]/[001]$ should appear as circular/elliptical in (b) $\mathbf{g} = (01\bar{1})$, respectively[82].

The average size and number densities of the defect clusters and dislocation loops are summarized in **Table 3**. Here, to make the comparison meaningful at a similar dpa range (7~10dpa), the damage peak (the dislocation band) region in the β-phase was not included for counting. The result shows quite contrasting radiation damage responses between α and β matrices: In the α-phase matrix, the size of the loops was about 2 to 3 nm both for *a/c*-loops and the number densities of ~$10^{23}$ m$^{-3}$. On the other hand, for the β-phase matrix, the number density of the $1/2\langle 111\rangle$ loops was significantly lower, only about 3~4×$10^{20}$ m$^{-3}$, and the average size was significantly larger (20~30 nm). So far studies on the radiation damage effects on β-titanium are quite limited. For the high-energy proton irradiation on a metastable β-phase alloy to 0.1 dpa, damage signatures were not observed at all [71]. As for the author's knowledge, this is the first observation of radiation-induced dislocation loops with finite-size in β-phase titanium.

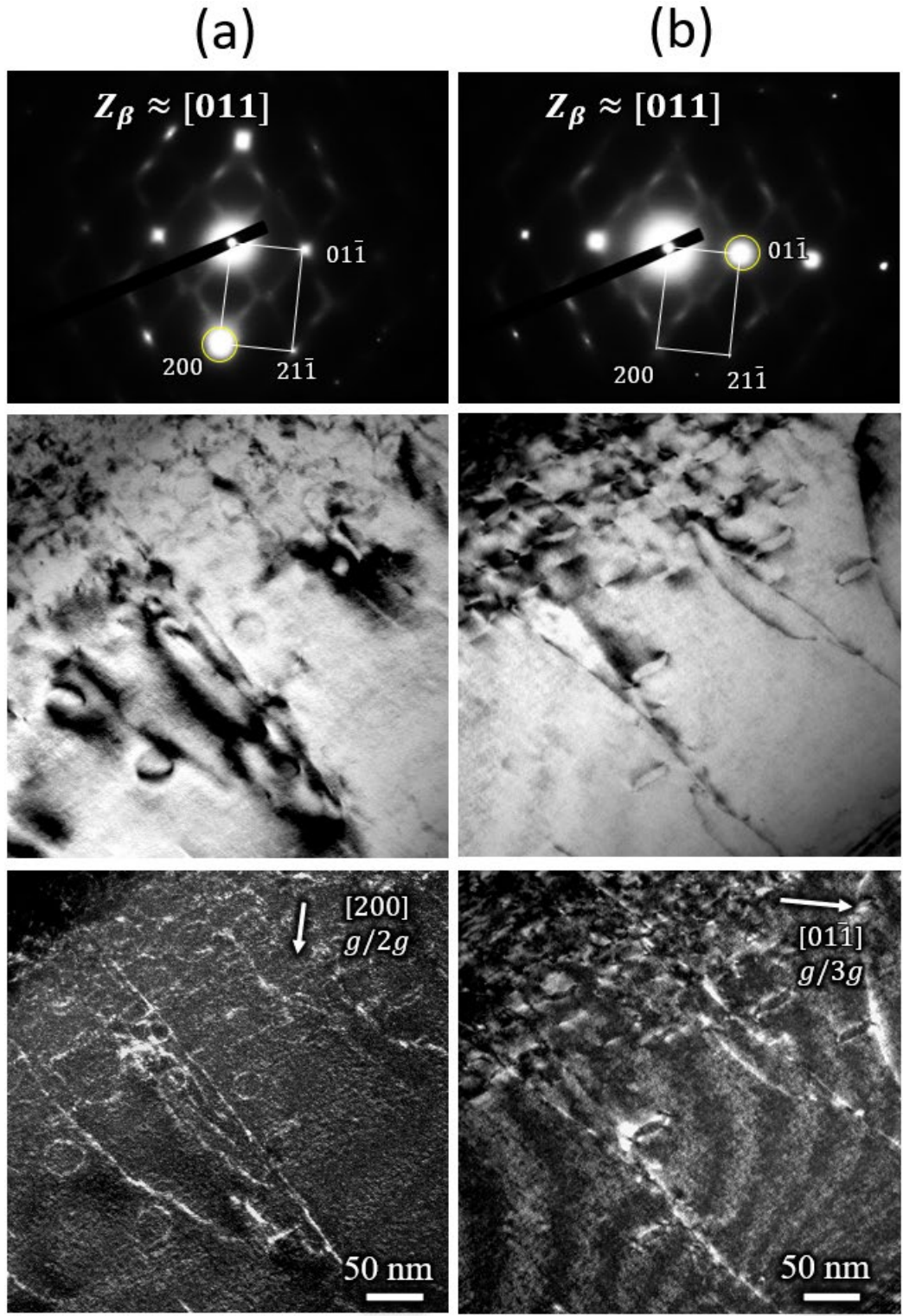

Fig. 11. SADP (top), BF (middle), and WBDF (bottom) images of β-phase matrix with $Z \approx [011]_\beta$ zone axis for an area with accumulated dose ranging 7 to 11 dpa under RT, increasing from bottom-right to top-left. The WBDF images were taken by selecting a diffraction spot indicated by a yellow cycle in each DP.

To correlate radiation damage signatures in microscopic observations to changes in the hardness and strength of polycrystalline metals, the dispersed barrier hardening model[83,84] has been widely employed so far. Geometrical consideration of dislocations in motion on a slip plane intersecting with randomly dispersed obstacles leads to an equation:

$$\Delta\sigma = \alpha \cdot MG \cdot b \cdot (Nd)^{1/2},$$

where $\Delta\sigma$ is the increase of the uniaxial tensile stress, $\alpha$ is the averaged barrier strength of the obstacles, $M$ is the Taylor factor (*i.e.* ratio of yield stress to critical resolved shear stress in uniaxial tension of polycrystalline materials), $G$ is the shear stress, $b$ is the magnitude of Burgers vector of dislocation, and $N, d$ are atomic density and diameter of the dispersed obstacles, respectively. For α-titanium, $M$ = 5.0[85,86], $G$ = 39.8 GPa, and $b = a_0$ = 2.95Å (length of the lattice vector in the basal plane). The barrier strength of the radiation-induced defect clusters was estimated for Zr to be $\alpha$ = 0.5, with assumptions of $M$ = 3.06 (valid for FCC and BCC crystals) and $b = \sqrt{5}a_0/3$ [74]. By substituting $M$ = 5.0 and $b = a_0$, the corrected obstacle strength can be obtained as $\alpha$ = 0.23[87], which is a comparable value to the BCC/FCC metals[74]. From the numbers given in **Table 3**, $(Nd)^{1/2}$ is calculated to be 1.8×$10^{-4}$ m$^{-1}$ for the observed *a*-loops, where the atomic density $N$ is multiplied by 1.5 to account for the variants invisible due to $g \cdot b = 0$. As a result, $\Delta\sigma$ is estimated to be 0.24 GPa, which is quite consistent with the observed increase of the strength, about 200 MPa. $(Nd)^{1/2}$ for the larger/fewer dislocations in the β-phase is calculated to be 3.1×$10^{-6}$ m$^{-1}$, *i.e*, more than factor 50 less than that for α-phase. The formation of the high-density defect clusters /dislocation loops in the dominant α-phase matrix is thus considered to be the dominant cause of the observed irradiation hardening. It is to be noted that the dispersed barrier hardening model is applicable mainly for more isotropic FCC and BCC polycrystals, where slip is the main cause of plastic deformation. In the case of α(HCP)-titanium with a small *c/a* value, the slip direction of the primary slip systems is only in the basal plane, and deformation twinning is the main deformation in the *c*-axis direction, whose radiation hardening effect should be modeled differently.

| α-phase, 7dpa | | | | β-phase, 7~10dpa | |
|---|---|---|---|---|---|
| \<a\> component loops | | \<c\> component loops | | 1/2\<111\> loops | |
| Size (nm) | Density (m$^{-3}$) | Size (nm) | Density (m$^{-3}$) | Size (nm) | Density (m$^{-3}$) |
| 2.3 | 9.4 × $10^{22}$ | 2.7 | 3.2 × $10^{22}$ | 23~32 | 2.7~4.5 × $10^{20}$ |

Table 3 Average size and number density of dislocation loops observed in each phase matrix of the Ti-64 specimen, with accumulated dose of 7 dpa for α phase / 7~10dpa for β phase under RT.

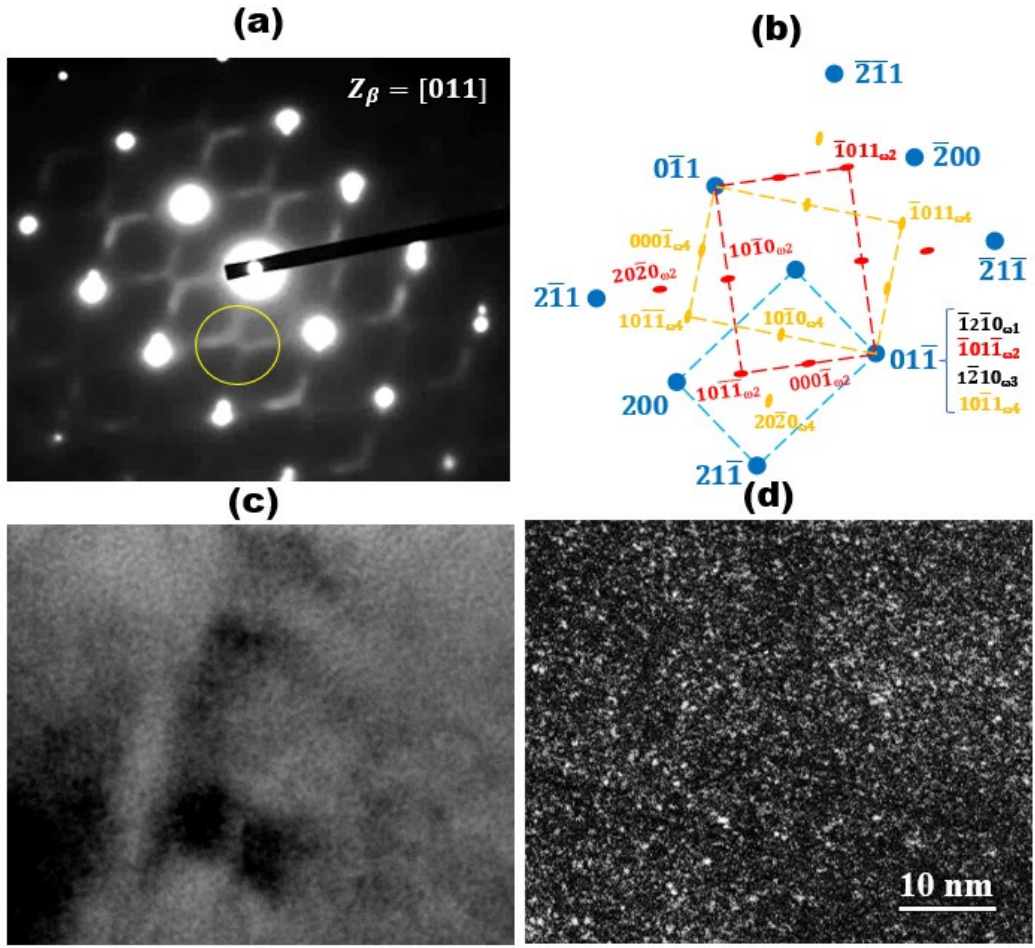

Fig. 12. High-magnification TEM images of β-phase matrix of the Ti-64 specimen irradiated to 11 dpa under RT, with Z-direction parallel to Z = [011]$_β$ zone axis: (a) SADP and (b) schematic with indices of b/w-phase reflections, (c) BF image, and (d) DF image of the area including diffuse streaks, as indicated by a yellow circle in the SADP.

3.3.2 TEM observation on ω-phase in β phase matrix

Fig. 12 is the SADP, BF, and DF micrographs of the β (BCC)-phase matrix with $Z = [011]_β$ zone axis in higher magnification. In (a)SADP between discrete spots of the mother β-phase (some of these spots are doubled due to the overlapping of secondary α platelets in the β-phase under Burger's coordinate relationship[47] with Z = $[01\bar{1}0]_α$), characteristic rectilinear diffuse streaks were identified clearly. By selecting an area including reflections of the ω-phase[30] as marked with a yellow circle, numerous dot contrasts of size less than 1 nm were identified in (d)DF image, which are the ω-phase precursors. The estimated number density, $10^{24}$ m$^{-3}$ or even more, is much higher than the observed defect densities in the α and β phase matrix. In the (c)BF image, a mottled background in a very fine scale was observed, which is also due to the high-density ω-phase precursors.

For high-energy proton irradiation of 0.25 dpa at <120°C, the diffuse streaks in the diffraction pattern of the β-phase matrix in the unirradiated state were transformed into discrete spots corresponding to the fully-developed ω phase precipitates[42]. To investigate dose dependency in the manifestation of the ω phase for current ion beam irradiation, a series of SADPs were taken at different depths of an identical β-phase matrix as collected in Fig. 13. Here, to avoid the influence of the matrices other than β-phase (such as platelet secondary α-phase), regions with DP showing single BCC matrix reflections were selected, where EDS on these regions show a high concentration of vanadium, the β-stabilizing element, 22.7±0.6 wt.% (low concentration of aluminum, 1.9±0.5 wt.%, α-stabilizing element). The contrast from the diffuse streaks was visible in all dose regions, whose pattern and intensity are quite stable and almost independent of the irradiation dose, which is in strong contrast to the proton irradiation result. At this moment we do not have an explanation for this difference, while it can be due to a difference in damage rate and irradiation temperature[88]: Compared to the current ion irradiation study, the irradiation rate for the proton beam irradiation was much lower, $3\times10^{-7}$ dpa/sec. According to an in-situ study of the diffraction pattern of ω-phase irradiated with 1 MeV electron beams in a high-voltage electron microscope, the diffuse scattering gradually changes from streaks to discrete spots at 450 K, but not at RT (Fig.6.13 of Ref.[26], and also Ref.[89]). For further investigation, it is worth repeating the ion irradiation study at higher temperatures.

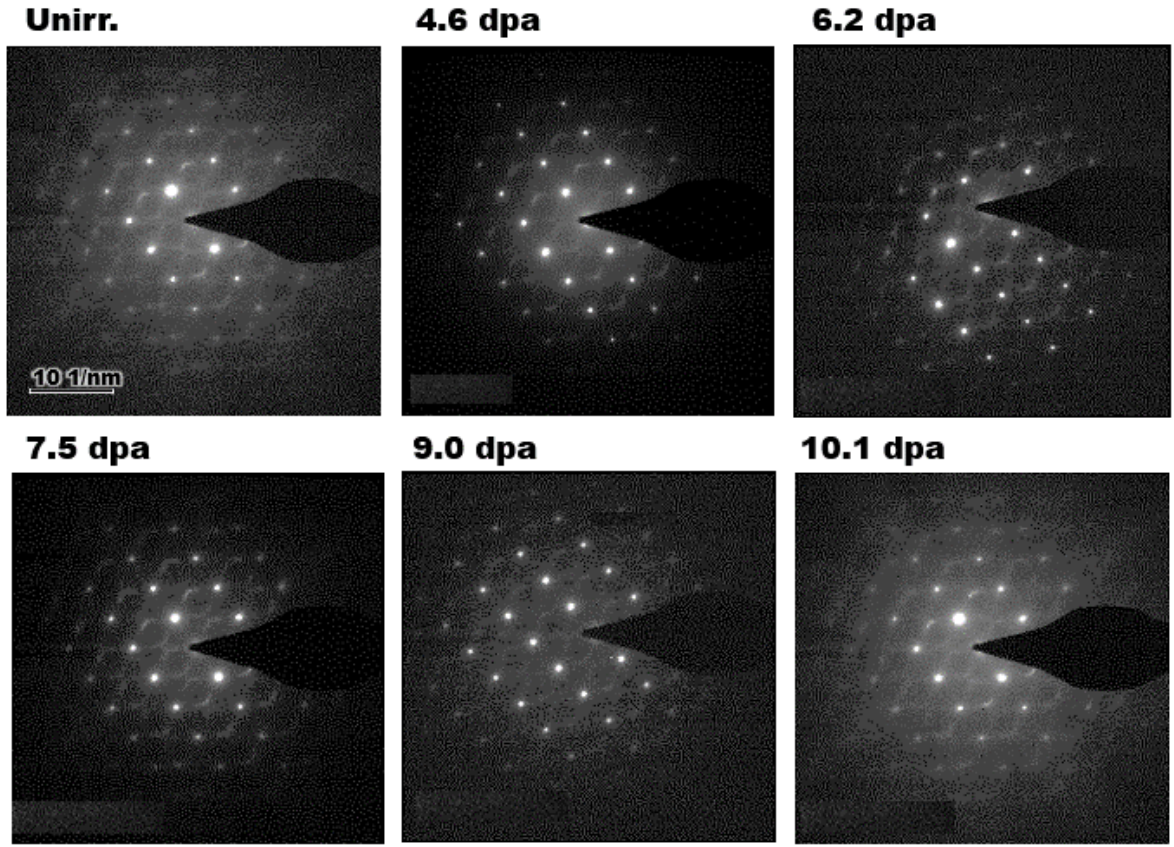

Fig. 13. Dose dependence of SADP, taken from an identical β-phase matrix at different depths from the surface of the specimen. The Z direction is parallel to the [011]$_β$ zone axis.

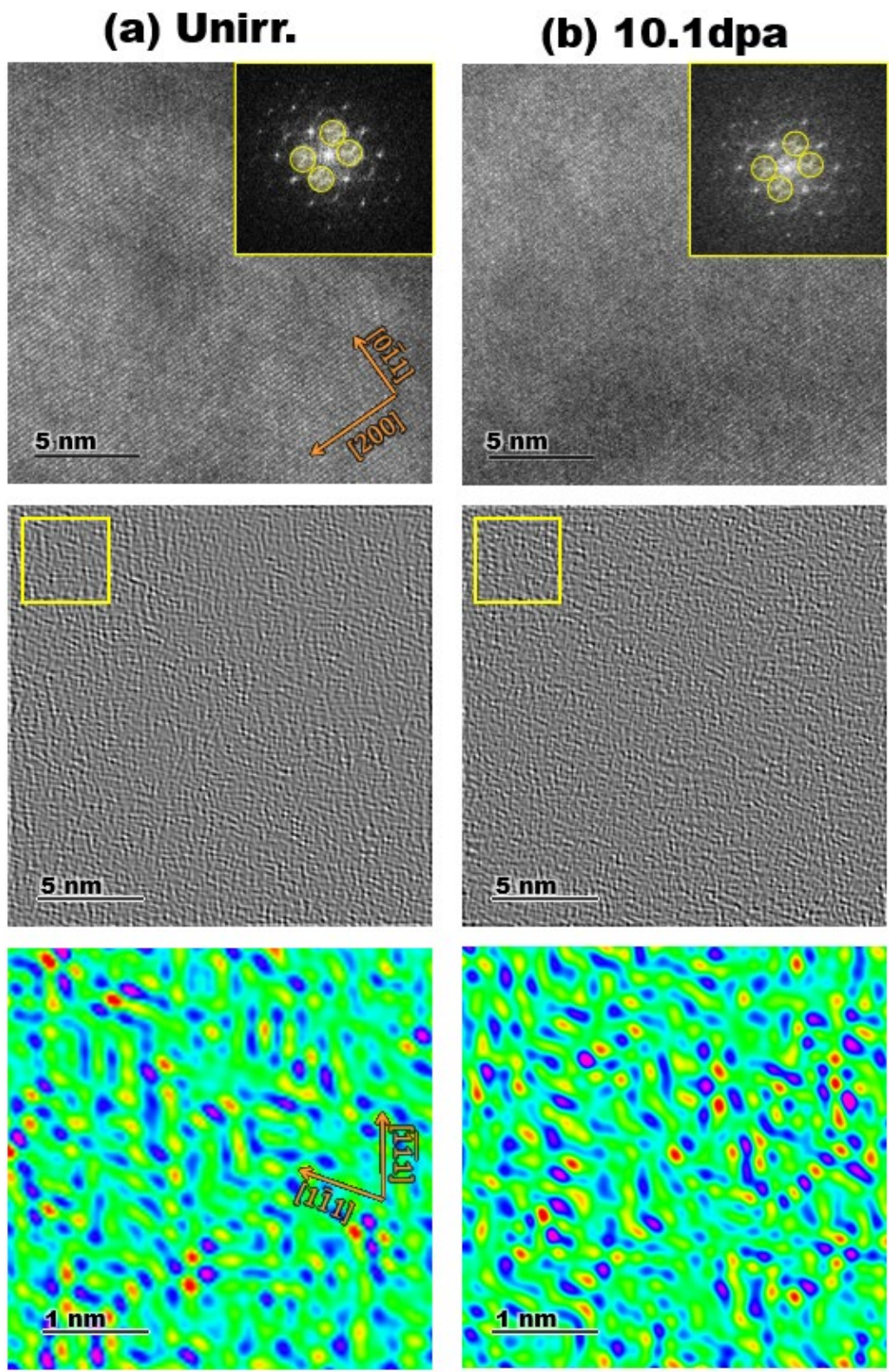

Fig. 14. HREM images of Ti-64 β-phase of the unirradiated (a) and irradiated (b) regions in Z= [011]$_β$ direction. (Top) the BF lattice image with the FFT as subset, (middle) the I-FFT image, utilizing area corresponding to the diffuse streaks shown as 4 yellow circles in the corresponding FFT, and (bottom) an enlarged view of the area indicated by the yellow box in the corresponding I-FFT image. Contrast is highlighted in rainbow colors (dark contrast in red/yellow, and light contrast in blue/purple).

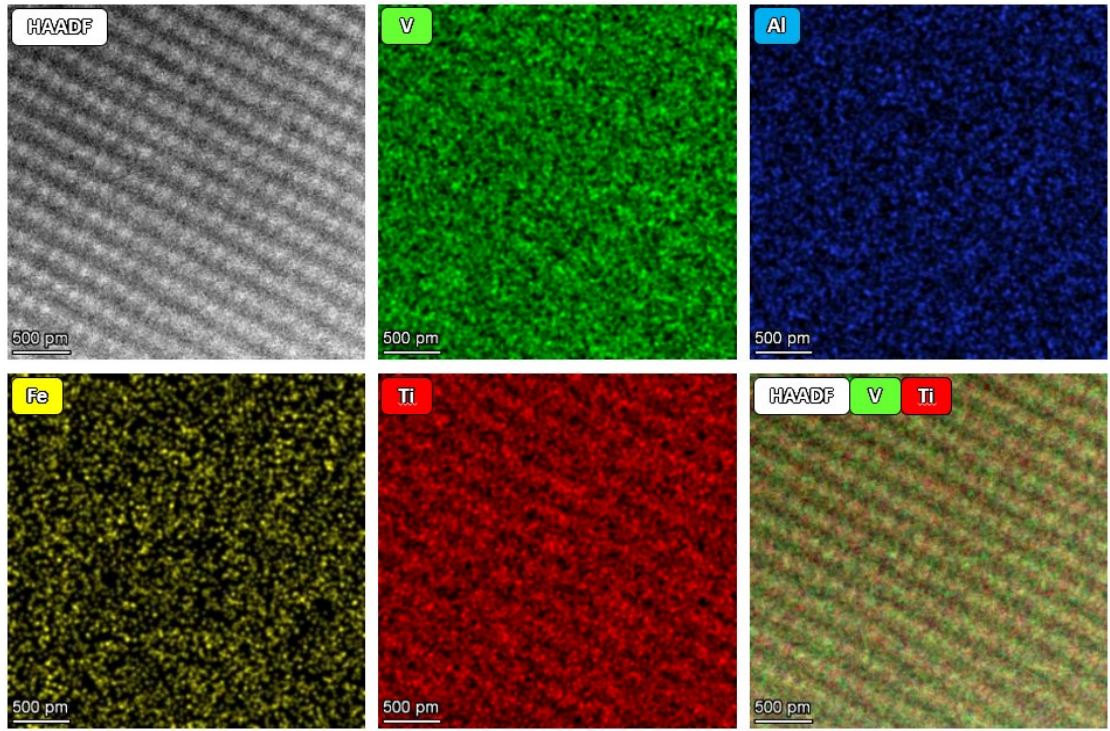

**Fig. 15. HAADF lattice image of a region close to the sample surface, overlaid with elemental mapping by EDS. The expected dose for the region was 4.6 dpa.**

To investigate the development of the ω phase structure in an atomic lattice scale, HREM analysis has been performed at both the un-irradiated and irradiated regions in the identical β-phase grain of the irradiated specimen, as shown in **Fig. 14**. Here, the irradiated region was taken at the depth from the incident surface corresponding to 10.1 dpa, while the non-irradiated region was taken at even greater depths where the ion beam couldn't reach to. In both lattice images (top), the structure of the β (BCC)-phase is recognized clearly, but cluster regions with distinct atomic displacements/collapse in a specific $\langle 111 \rangle_\beta$ direction, characteristic of the well-developed ω phase, cannot be identified. The sample thickness in these regions measured with Electron Energy Loss Spectroscopy (EELS) was about 200 nm, and the atomic displacements in nano-scale regions might be averaged out and difficult to see. The fast Fourier transform (FFT) analysis, shown in the subset of each lattice image, exhibited diffuse streaks in addition to the mother β-phase spots, as observed in the electron diffraction in the TEM analysis (**Fig. 12**a and **Fig. 13**). The Inverse fast Fourier transform (I-FFT) analysis of the region covering the diffuse streaks around the central transmission wave, indicated by four yellow circles in the corresponding FFT image is shown in the middle figures. It reveals a distinctive sub-nanometer-sized lattice disorder with local fluctuations, but continuous rather than discrete, and almost homogeneous within the mother β(BCC)-phase lattice. In the zoom-up of the I-FFT image (bottom), the sequence of atoms can be identified clearly as contrasts with linear and/or row orientations parallel to the $\langle 111 \rangle_\beta$ directions, appearing as lines of about 1 nm-long, or even smaller rectangular clusters. This homogeneous fine structure in the β-phase is identified as the ω-precursor. Since the lattice disorder was observed in almost similar pattern and size between un-irradiated and irradiated regions, it is considered that the ω-precursors formed during the fabrication process stayed unchanged even at ~10 dpa.

Fig. 15 shows a HAADF lattice image of a region close to the sample surface with less thickness of about 150 nm (irradiation dose of 4.6 dpa), overlaid with elemental mappings by EDS. In the HAADF lattice image (top-left), no clear atomic displacements are observed in the mother BCC lattice, and the cluster-like segregation of the Ti-enriched region, which is reported in Ref.[34] as characteristic of the developed ω phase, does not seem to be visible. It is consistent with the nature of the ω-precursor, being formed as a purely displacive process without compositional fluctuation.

## 3.4 Possible reasons for contrasting irradiation response between α and β phases

Through TEM observation, we have confirmed the significantly low dislocation density and the absence of phase transformation in the β-phase matrix in comparison to the α-phase. In a recent primary knock-on atom (PKA) simulation on Ti-64, it is shown that the displacement threshold energy for α-phase (66 eV) is higher than that for β-phase (46 eV), and number of Frenkel pairs produced by the 40 keV PKA in β-phase is about factor 2 to 4 larger than that in α-phase[90]. Thus the difference in radiation response of the two phases is likely to arise from longer-term defect evolution. In this context, we investigate possible two causes, *i.e.* (1) the strong sink effect expected for the distinctive sub-nanometer-sized homogeneous lattice disorder as observed in HREM, and /or (2) the anomalous point defect recombination induced by the high mobility of vacancies. Both of these effects originated from the metastable ω-phase precursors specifically formed in the β(BCC) phase of group-4 transition metals.

### 3.4.1 Sink effect caused by the sub-nanometer-sized homogeneous lattice disorder

Irradiation damage "*sink sites*" refer to crystalline structures that act to promote the recombination of point defects produced by irradiation and mitigate their concentration, which include (1) dislocations, (2) grain boundaries and (3) particle precipitates[72,91]. With respect to (3), nano-structured materials[92] such as Oxide-Dispersion-Strengthened (ODS) alloys[93] show superior irradiation-resistant properties, owing to the large free surface to volume ratio of interphase boundaries, compared to conventional materials having grains larger than micrometer scale. The "*sink-strength*" of the particle precipitates $S_p$ [m$^{-2}$] can be approximated as:

$$S_p = 2\pi N_p d_p,$$

where $N_p$ [m$^{-3}$] is a number density of particle precipitates with diameter $d_p$ [m]. Here, the equation is considered valid for the "incoherent" particles, which have a crystal structure different from the mother matrix without any orientational relationships. Meanwhile, since oxide nano-particles in the ODS alloys are found to be generally coherent with the mother matrix[94], we assume the ω precursors, sub-nanometer-sized lattice disorder with specific orientational relationships with the mother matrix, may also possibly act as the point-defect sink site. Although it is not straightforward to define particle size and density for the homogeneous lattice disorder, the DF image by selecting diffuse streaks of the ω-precursor (Fig.12d) gives particle size and density as $d_p = 0.6 nm$ and $N_p = 1 \times 10^{25} m^{-3}$, respectively. From these numbers, we can deduce the sink strength as $S_p = 3.8 \times 10^{16} m^{-2}$, which is larger by one order than that for the ODS Fe-Cr alloys, $2.7\sim6.4 \times 10^{15} m^{-2}$ [91]. The mechanism of the lattice structure corresponding to the ω-precursor acting as point defect sink is worth investigating further.

### 3.4.2 Point-defect recombination due to anomalous self-diffusion in β-Ti

The anomalous self-diffusion in the β phase of the group 4 transition metals[95] is worth to be investigated. In typical pure metals including α-titanium, the self-diffusion coefficient follows the Arrhenius law:

$$D_\alpha = D_0 \exp\left(-\frac{Q}{k_B T}\right),$$

where $D_0$ is a pre-exponential factor (for α-Ti, $D_0 = 1.35 \times 10^{-3} m^2/s$), Q is the activation energy (3.14 eV), $k_B$ is Boltzmann constant (= 8.617 eV/K), and *T* is the absolute temperature, respectively[96]. Here, the activation energy *Q* can be approximated to be 34×$T_m$ for typical metals, where $T_m$ is the absolute temperature of melting. In case *D* is expressed in a logarithmic scale as a function of $T_m/T$, all typical metals obey about the same straight-line function, because self-diffusion in typical solid metals occurs by the common "*single vacancy*" mechanism, *i.e.* atoms jump to the nearest site via a single vacancy due to thermal excitation. However, the self-diffusion coefficient of the β(BCC)-phase group-4 transition metals (Ti/Zr/Hf) is not only larger by some orders than those for the normal metals, but also shows an upward bending with enhanced diffusion at low temperatures, which is expressed by:

$$D_\beta = D_0 exp\left(-\frac{Q}{k_B T}\right) exp\left(\frac{\alpha}{k_B T^2}\right),$$

where $D_0 = 3.53 \times 10^{-4} m^2/s$, $Q = 3.40 eV$, and $\alpha = 1{,}335\ eV \cdot K$ for β-titanium[96]. At the β-to-α transus temperature (T = 1,155K), $D_\beta / D_\alpha$ is calculated to be $2.1 \times 10^3$, thus self-diffusion in the β-phase is larger by more than three orders than in the α-phase, which may correspond to the diffusion of α-phase about at +400 K higher temperature. Note that $D_\beta / D_\alpha$ could be even larger at the RT.

This anomalous self-diffusion near the transus can be reproduced quantitatively by a model based on the identity of the activated complex of the diffusion and the basal plane of the metastable low-temperature ω phase[97,98]: in the β-phase of group 4 transition metals, as the temperature decreases, the atoms in the body-center position are displaced onto the (111) plane, and become activated in its center of the triangle. With this atomic configuration, being identical to the ω-phase precursor, it is easier to exchange positions with the vacancies in the neighboring BCC position, and thus diffusion is enhanced near the transus. In the present study, the fact that irradiation-induced defect clusters were merely observed in the β-phase, could be explained by the same mechanism, *i.e.* the lattice disorder corresponding to the ω-phase precursor preserves several orders higher mobility of the vacancies, which facilitates anomalously fast recombination of the radiation-induced point defects, and as a result radiation damage microstructure is *wiped out* even at the room temperature, as if the irradiation occurs at much higher temperature regime. This anomalous point defect recombination hypothesis is worth further investigating in quantitative manner, by means of a molecular dynamics simulation *etc.*

# 4. Conclusions

Aiming to investigate the radiation damage effect on the Ti-64 alloy utilized as accelerator beam window material, a series of irradiation experiments have been conducted with a 2.8 MeV-$Fe^{2+}$ ion beam. The radiation hardening and microstructural changes are studied in detail up to 11dpa by the nano-indentation hardness measurements and TEM/HREM observations. Here, the utmost attention was set on the irradiation-induced ω-phase transformation in the β-phase reported in a former high-energy proton beam irradiation, as the possible source of the embrittlement observed for this alloy. The results are summarized as follows:

- From the nano-indentation hardness measurements, the magnitude of irradiation hardening was 1.4~1.5 GPa in Ti-64 after the irradiation over 1 dpa, which stays almost unchanged up to 11 dpa. TEM observation identified the formation of both *a*- and *c*-loops with the size of 2~3 nm and density of ~$10^{23}$ $m^{-3}$ in the dominant α-phase matrix, which is considered the main contributor to the observed radiation hardening, as evaluated by the dispersed barrier hardening model.
- For the β-phase matrix in the region with damage less than 10 dpa, the larger dislocation loops were identified with a significantly lower density than the α-phase. Instead, characteristic diffuse streaks in the diffraction pattern and fine precipitations in the dark field image were identified with an average size of less than 1 nm and extremely high number density, which stayed stable after the irradiation. They are the ω-phase precursors commonly observed in the metastable β-phase Ti-alloys, but so far merely reported for the dual-phase Ti-64 alloy.
- FFT/I-FFT analysis of the HREM for the streaks revealed a sub-nanometer-sized lattice disorder with local fluctuations, continuous rather than discrete, and almost homogeneous within the mother β-phase. This distinctive lattice structure was stable against irradiation and remained unchanged even at 10 dpa.
- The fairly low density of the dislocation loops in the β-phase can be due to either (1) the strong sink effects around the ω-phase precursors manifesting as the homogeneous sub-nanometer-sized lattice disorder, or (2) higher mobility of vacancies boots up the recombination to interstitials and thus wiped out the radiation damage microstructure in the β-phase. This effect could be referred to as *anomalous point defect recombination*. In both cases, the ω-phase precursors, which are specifically formed in the β(BCC)-phase of group-4 transition metals play an essential contribution.

In beam window application, it is desirable to find alternative grades of titanium alloys that can maintain not only mechanical strength but also ductility in the high dose range. To this end, it is quite intriguing that the β-phase of titanium alloys may preserve higher radiation damage tolerance up to 10 dpa. We have already reported that the metastable β phase Ti-15V-3Cr-3Al-3Sn (Ti-15-3) alloy used as a beam monitor at J-PARC does not undergo irradiation hardening up to 0.1 dpa[71]. TEM microstructural studies yielded similar results to the Ti-64 β-phase of the present study: no defect clusters were observed, and instead dense ω-phase precursors remained unchanged after irradiation. The Ti-15-3 specimens have also undergone ion beam irradiation at much higher dose regions, and the results will be reported in a separate publication[99].

## Acknowledgements

This study was supported by the U.S.-Japan Science and Technology Cooperation Program in High Energy Physics, and Grant-in-Aid for Scientific Research (A) (21H04480 and 21H04668). This study is a collaborative research project at Nuclear Professional School, School of Engineering, The University of Tokyo. A part of this work was conducted at the Hokkaido University as a program of "Nanotechnology Platform" and "ARIM" of the Ministry of Education, Culture, Sports, Science and Technology (MEXT), Japan. A part of this work was also supported by Electron Microscopy Unit, National Institute for Materials Science (NIMS), Japan. The authors would like to thank Mr. Takao Omata (Univ. Tokyo) for his support of the HIT facility operation, Mr. Ryo Ohta (Hokkaido Univ.) for conducting HREM observations, Dr. Minghui Song and Dr. Yoshiko Nakayama (NIMS) for part of TEM observations. The authors thank to high-power target group of Rutherford Appleton Laboratory, Dr. Christopher J. Densham and his colleagues, for the provision of the beam window drawing ([Fig.1](b)).

*Statement: During the preparation of this work the corresponding author used DeepL in order to improve readability and language. After using this tool, the author reviewed and edited the content as needed and takes full responsibility for the content of the publication.*